\documentclass[twocolumn,showpacs,preprintnumbers,amsmath,amssymb]{revtex4}
\usepackage{color}

\def\pra{{Phys.~Rev.~A}}

\def\jcp{{J.~Chem.~Phys.}}

\usepackage{isomath}

\usepackage{upgreek} 

\usepackage{graphicx}
\usepackage{dcolumn}
\usepackage{bm}
\usepackage{rotating}
\usepackage{amssymb}
\usepackage[latin1]{inputenc}
\usepackage{natbib}
\usepackage{epstopdf}
\graphicspath{{./figures/}}

\begin{document}

\title{Rovibrational cooling of molecules by optical pumping}

\author{I. Manai}
\affiliation{Laboratoire Aimé Cotton, CNRS, Université Paris-Sud, ENS Cachan, 11, 91405 Orsay, France}

\author{R. Horchani }
\affiliation{Laboratoire Aimé Cotton, CNRS, Université Paris-Sud, ENS Cachan, 11, 91405 Orsay, France}

\author{A. Fioretti}
\affiliation{Dipartimento di Fisica, Università di Pisa and CNISM, Largo Pontecorvo 3, 56127 Pisa, Italy}

\author{M. Allegrini}
\affiliation{Dipartimento di Fisica, Università di Pisa and CNISM, Largo Pontecorvo 3, 56127 Pisa, Italy}

\author{H. Lignier}
\affiliation{Laboratoire Aimé Cotton, CNRS, Université Paris-Sud, ENS Cachan, 11, 91405 Orsay, France}

\author{P. Pillet}
\affiliation{Laboratoire Aimé Cotton, CNRS, Université Paris-Sud, ENS Cachan, 11, 91405 Orsay, France}

\author{D. Comparat}
\affiliation{Laboratoire Aimé Cotton, CNRS, Université Paris-Sud, ENS Cachan, 11, 91405 Orsay, France}

\date{\today}

\begin{abstract}
We demonstrate rotational and vibrational cooling of cesium dimers by optical pumping techniques. We use two laser sources exciting all the populated rovibrational states, except a target state that thus behaves like a dark state where molecules pile up thanks to absorption-spontaneous emission cycles. We are able to accumulate photoassociated cold Cs$_{2}$ molecules in their absolute ground state ($v=0,J=0$) with up to $40\%$ efficiency. Given its simplicity, the method could be extended to other molecules and molecular beams. It also opens up general perspectives in laser cooling the external degrees of freedom of molecules.
\end{abstract}
\pacs{}

\maketitle

It is commonly admitted that optical manipulation of molecules comes up against strong limitations. The difficulty originates from the large number of internal states accessible by spontaneous emission events \cite{2008_PRL_Ye_direct_cooling}. However, a few recent experiments succeeded to implement molecular optical pumping: for example, the vibrational cooling of Cs$_2$ \cite{MatthieuViteau07112008}, the optical cooling of SrF \cite{shuman2010laser}, or the rotational cooling of molecular ions \cite{2010NatPhDrewsen_rotational_cooling}. These methods, like the one we present in this article, are fundamentally different from coherent optical manipulations, such as STIRAP, that affect the population of a single quantum state \cite{JohannDanzl07102008,2010JinPRL_HyperfinePreparation}. 

Optical cooling of both the vibration and the rotation of Cs$_2$ molecules is complicated because these two degrees of freedom can not be manipulated independently. The pumping of one of them tends to impair the other one, i.e. the vibrational pumping is likely to modify the rotational quantum number, just as the rotational pumping is likely to modify the vibrational quantum number. As a consequence, a global rovibrational cooling can only be achieved through an interplay between both processes.
Our vibrational pumping, already demonstrated in \cite{MatthieuViteau07112008,2009_JMOp_Cooling,2009NJPh...11e5037S,2009PhRvA..80e1401S,2012_PRA_Horchani_Conversion}, makes use of a broadband laser whose spectrum has been specifically shaped to excite all the vibrational levels but one where molecules accumulate. The frequency resolution is limited to $\sim 0.1$ cm$^{-1}$. For rovibrational pumping, a requirement is the control of the light spectrum with a resolution on the order of magnitude of the rotational constants \cite{2009_JMOp_Cooling,2010MolPh.108..795S}. As the Cs$_2$ molecules have a rotational constant of about $0.01$ cm$^{-1}$, which is exceptionally small compared with the great majority of diatomic molecules, we have designed a method improving the spectral resolution of molecular optical pumping with respect to the grating technique. It consists in scanning a narrow-band laser diode in appropriate regions of the rotational spectrum, which allows one to modify the population distribution among the rotational levels characterized by their $J$ quantum numbers. 

Our experimental setup is based on a caesium magneto-optical trap (MOT). All the manipulations achieved on this source last a total of $100$ ms. During the first 20 ms, a cw Ti:Sa laser is focused on the MOT to form molecules in the electronic ground state by photoassociation (PA). The PA scheme, detailed in Ref. \cite{2011PCCP...1318910L}, consists in exciting a $0^-_g (6s + 6p_{1/2})(v'=26,J')$ rovibrational level, where $v'$ is the vibrational quantum number counted from the dissociation limit, and $J'$ the rotational quantum number. The rotational and vibrational cooling lasers are switched on at the same time as the PA laser, but are respectively applied for $25$ ms and $28$ ms. The vibrational populations are then probed by converting molecules into ions via a resonant enhanced 2 photon ionization scheme (RE2PI). For this purpose, a pulsed dye laser is turned on $1$ ms after switching off the vibrational cooling laser. It allows us to scan the vibrational transitions between the X$^1\Upsigma_{\rm g}^{+}$ state and the C$^1\Uppi_u$ or D$^1\Upsigma^+_u$ intermediate states \cite{2011PCCP...1318910L}. Cs$_2^+$ ions are finally detected by micro-channel plates. To access the rotational distributions, we insert a 50 $\mu$s pulse of a narrow band laser $0.5$ ms before the application of the ionizing pulsed laser.

\begin{figure}[htbp]
\begin{center}
\includegraphics*[angle=-90,width=1\columnwidth]{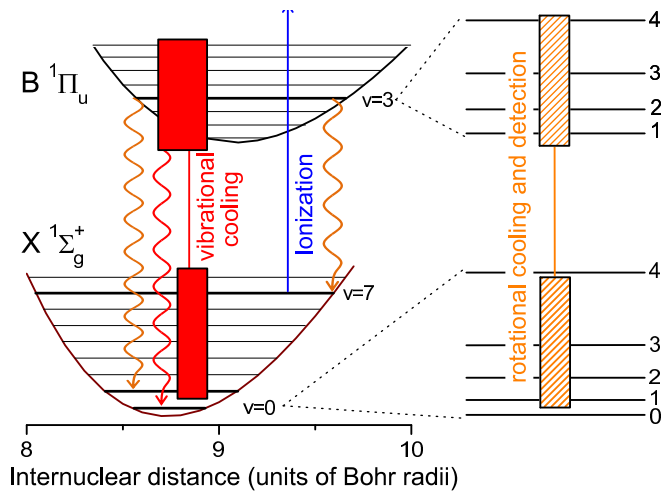}
\caption{(Color online) Transitions used to cool and detect the vibration and rotation of Cs$_{2}$ molecules. The (red) filled and (orange) hatched rectangles cover the vibrational levels affected by the broadband laser and the rotational levels excited by the two swept narrow band lasers used for rotational cooling and detection. The upper and lower numbers on the right hand side are $J'$ and $J$ values, respectively.  The (blue) straight arrow represents the RE2PI laser ionizing molecules through the C or D states (not shown). The wavy arrows indicate the spontaneous emission processes.}
\label{fig:figure1}.
\end{center}
\end{figure}

Vibrational cooling is performed according to Ref. \cite{MatthieuViteau07112008}, i.e. with a femtosecond laser ($200$ mW final power, $120$ fs pulse duration, $1$ mm$^2$ beam size) tuned to rovibrational transitions between the ground state and the B$^1\Uppi_u$ excited state as shown in Fig. \ref{fig:figure1}. This allows us to accumulate molecules in $v=0$. It is conceivable to extend the shaping technique of the femtosecond laser to rotational cooling provided that the molecular species under study has larger rotational constants \cite{2011PCCP...1318825L,2009_JMOp_Cooling} than the laser shaping resolution. We stress the fact that this requirement is not fulfilled in the case of Cs$_2$.

To manipulate the rotational populations of Cs$_2$, we employ a cw DFB diode laser ($\sim 2$ MHz linewidth, $\sim 6$ mW power, $3.5$ mm$^2$ beam size) scanned across the B$^{1}\Uppi_{\rm u}(v_B=3,J_B)\leftarrow $X$^{1}\Upsigma_{\rm g}^{+}(v=0,J)$ transitions though the diode current (see right part of Fig. \ref{fig:figure1}). The choice of these transitions originates from the wavelength accessibility of the diode laser. We can explore the $P$, $Q$, or $R$ rotational branches, schematically shown in Fig. \ref{fig:figure3}, and thus change the $J$ value as proposed in \cite{1996_JCP_BahnsLaserCooling}. The trend in the evolution of $J$ is governed by the branch selected at the excitation.

\begin{figure}[htbp]
\begin{center}
\includegraphics*[width=1\columnwidth]{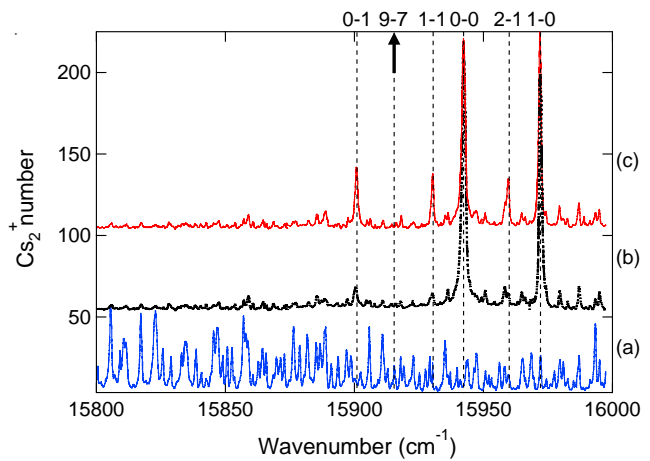}

\caption{(color online) Three vibrational spectra obtained by photoionization (RE2PI). (a) When only PA $0^-_g (6s + 6p_{1/2})(v'=26,J'=2)$ is applied, the spectrum reveals that molecules are mainly stabilized in the X state \cite{2011PCCP...1318910L}. (b) the application of the vibrational cooling provokes the emergence of a few intense lines while the intensity of the other lines is reduced. This proves the accumulation of molecules in $v=0$. (c) rotational and vibrational cooling to ($v=0,J=1$) leads to a spectrum similar to (b), but some vibrational levels $v\neq0$ are more populated than in (b). The number pairs above indicate the wavenumber of some $v_C-v$ transitions. The wavenumber of the $9-7$ transition, indicated by an arrow, is used for rotational spectroscopy. }
\label{fig:figure2}
\end{center}
\end{figure}

\begin{figure}[htbp]
\begin{center}
\includegraphics*[width=1\columnwidth]{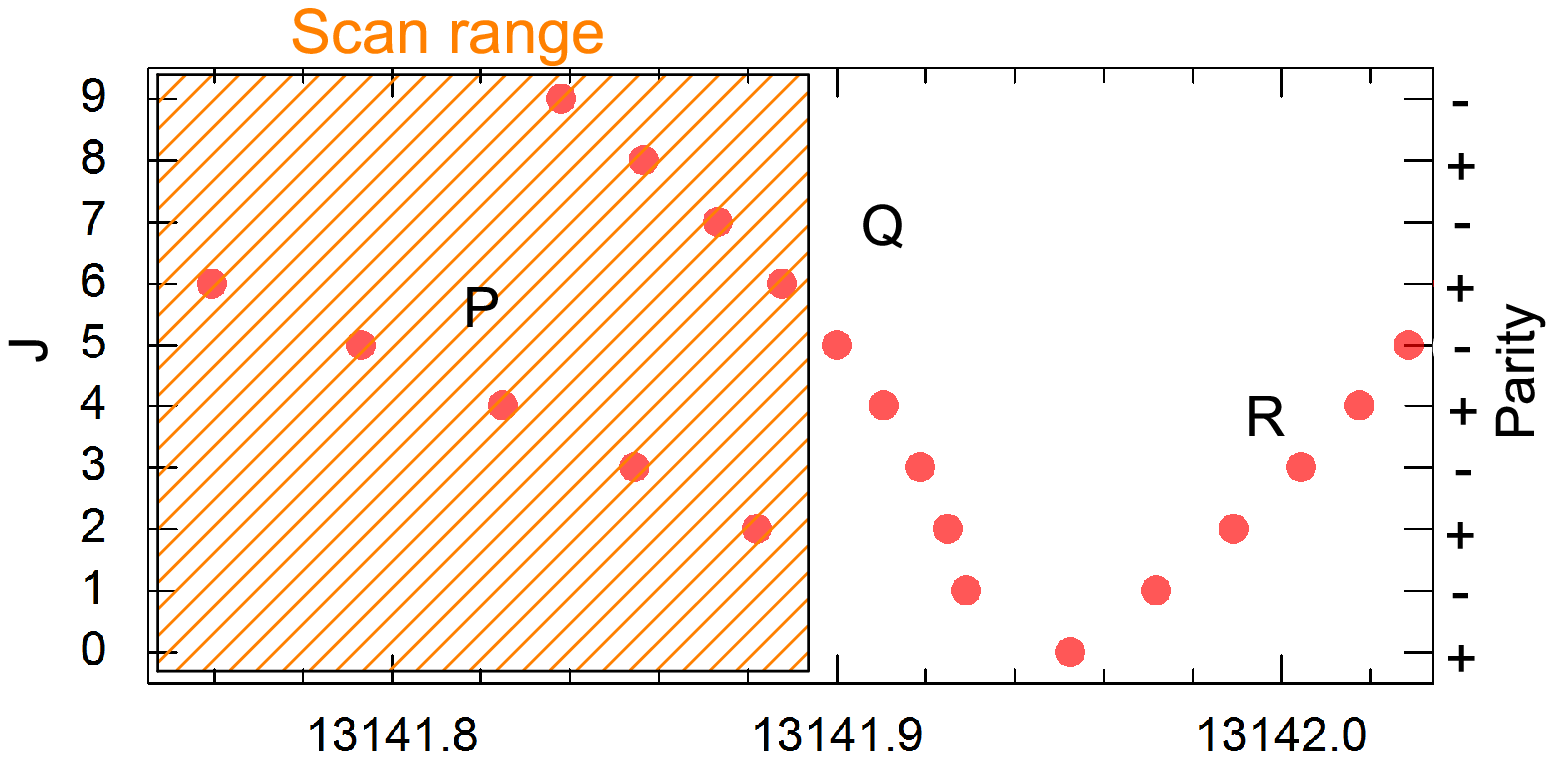}
\includegraphics*[width=1\columnwidth]{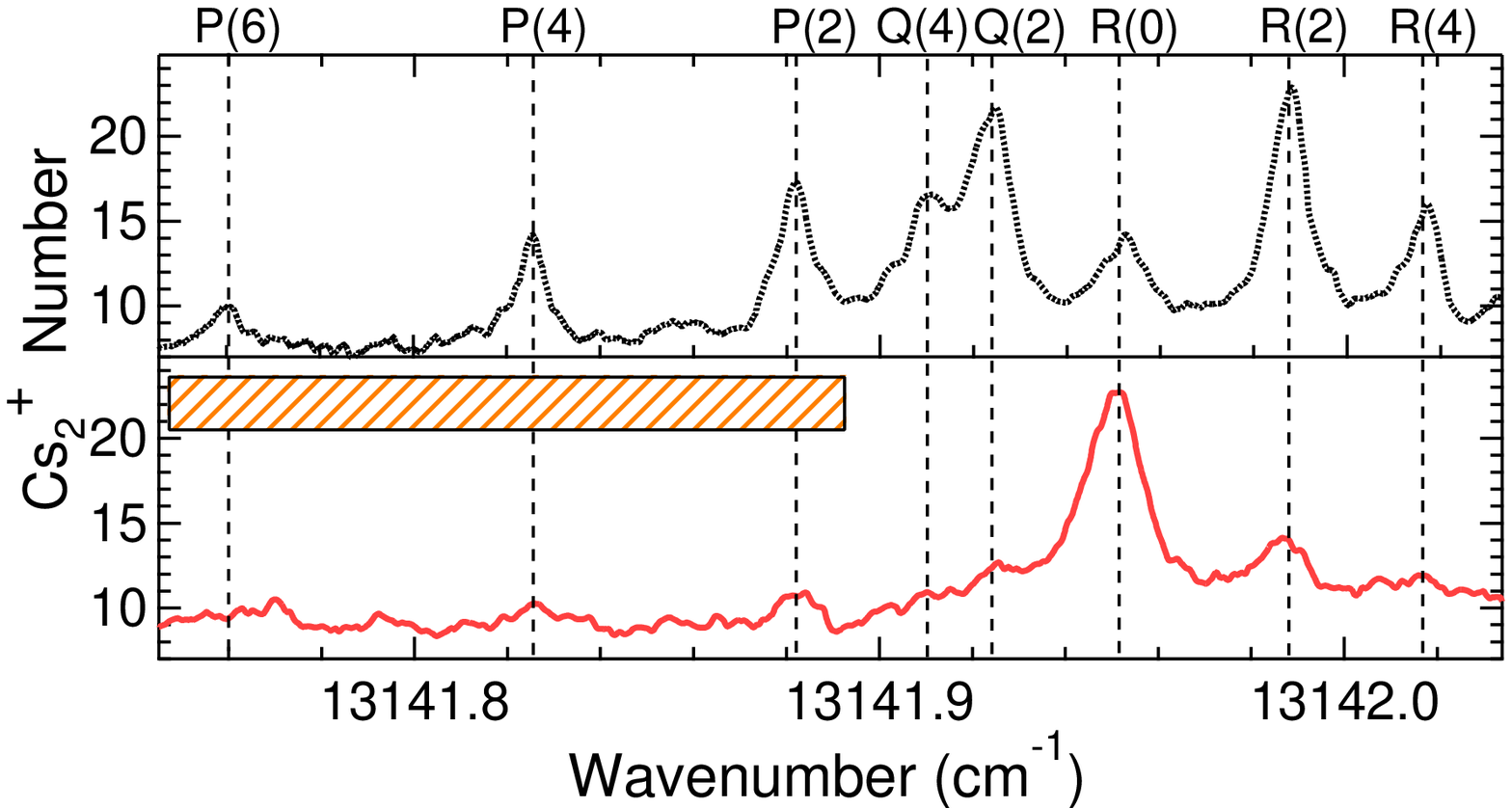}

\caption{(color online) Upper Panel: Fortrat diagram showing the transition wavenumbers from the $(v=0,0\leq J \leq 9)$ rotational levels of the X state. The $P$, $Q$, $R$ branches respectively correspond to the  $J_B-J=-1,0,1$ allowed transitions. The parity of the $J$ state is also indicated on the right hand side. The spectrum covered by the rotational cooling laser is shown by the hatched area. Lower panel: Comparison of rotational spectra obtained with vibrational pumping (upper black dotted line) and rovibrational pumping (lower red solid line). The narrow band laser used for rotational cooling is swept over the P(6), P(4) and P(2) transitions with repetitive $100$ $\mu$s frequency ramps. The rotational cooling efficiency is $\sim 40$\%.}
\label{fig:figure3}
\end{center}
\end{figure}

\begin{figure}[htbp]
\begin{center}

\includegraphics*[width=1\columnwidth]{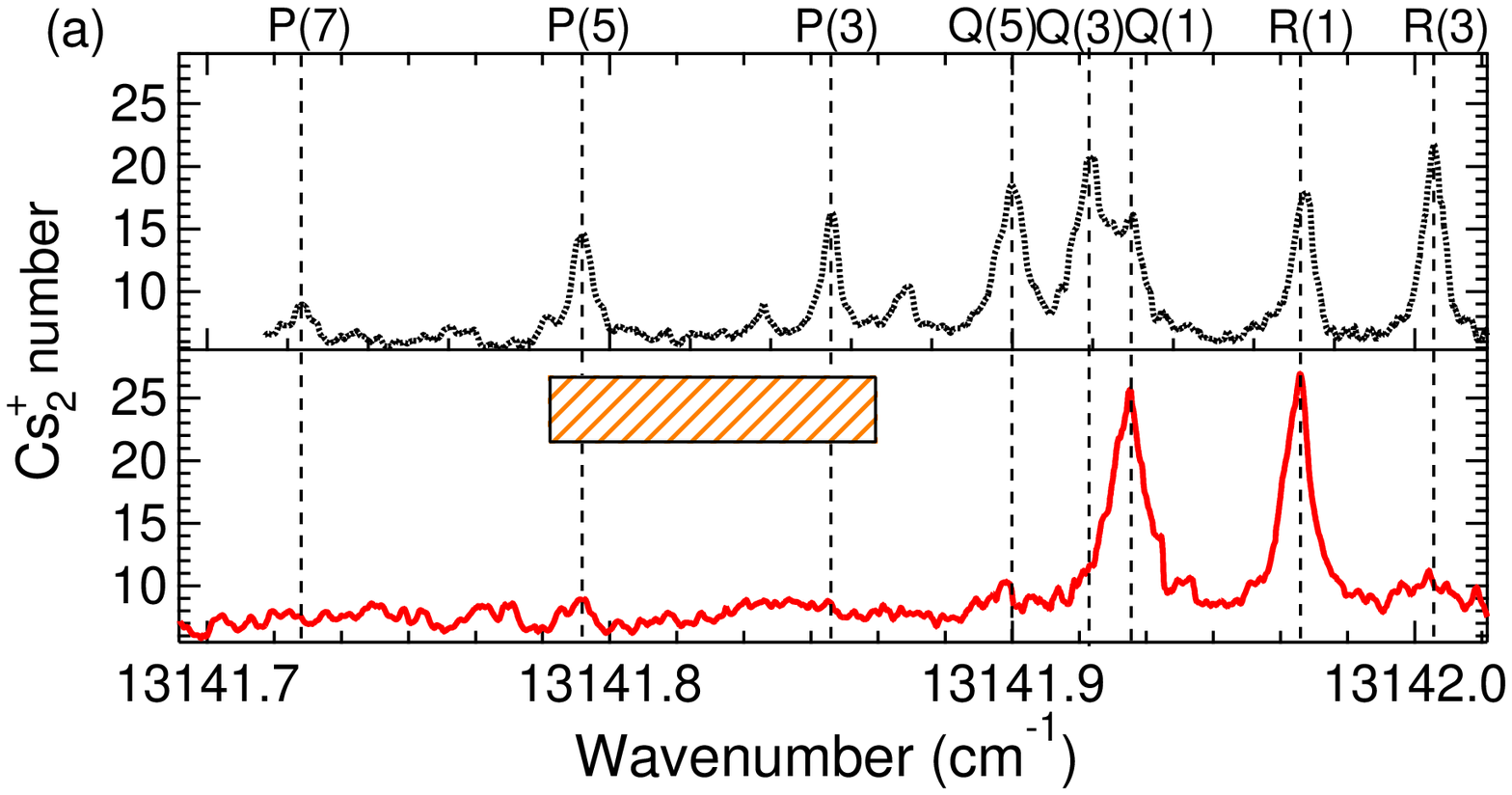}
\includegraphics*[width=1\columnwidth]{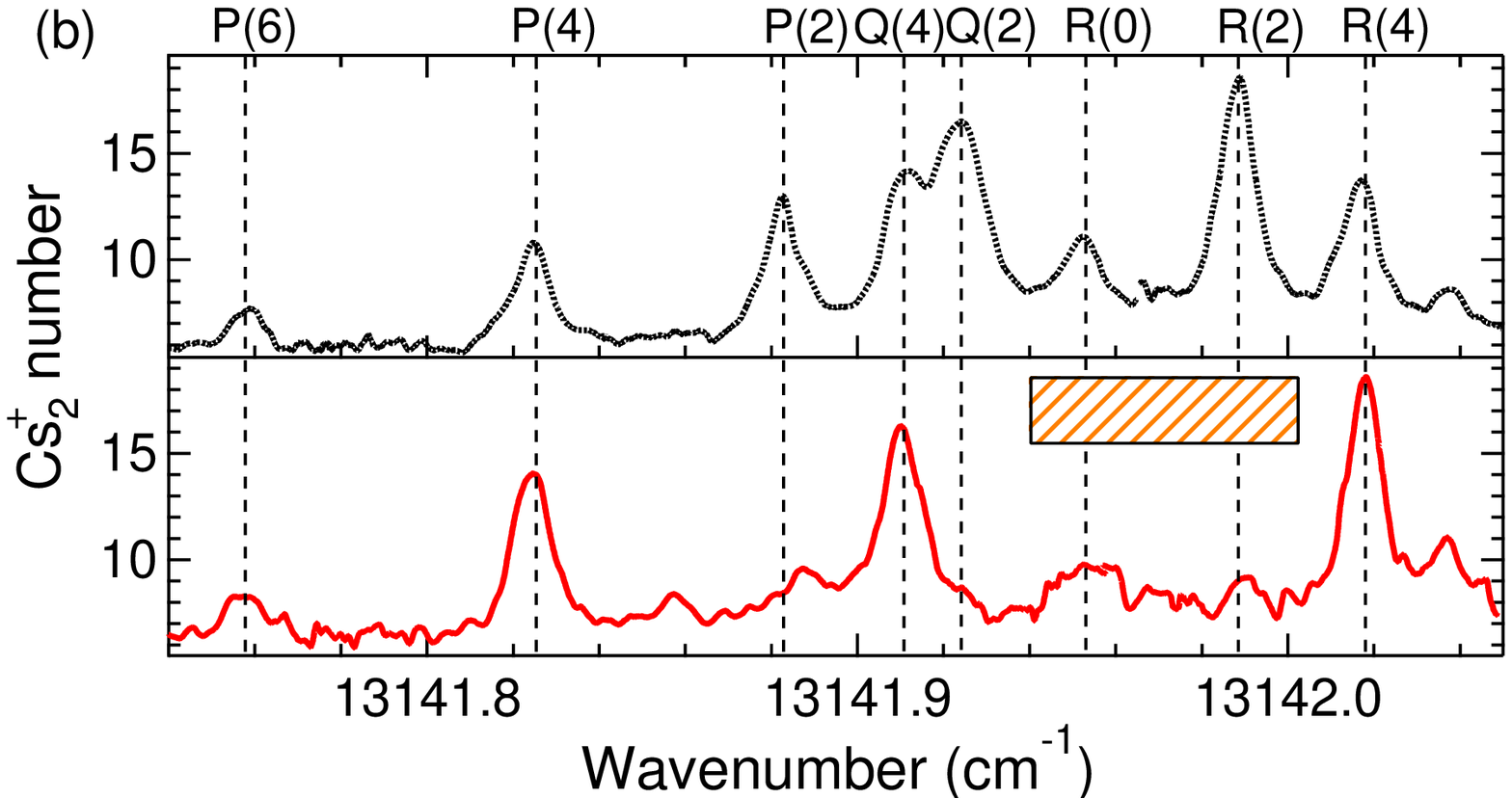}

\caption{(color online) Rotational spectra demonstrating the rovibrational pumping toward $J=1$ (a) and $J=4$ (b) when the narrow band laser is swept in the frequency range indicated by the hatched areas. For comparison, the spectra resulting from vibrational pumping (dotted line) are displayed above the spectra resulting from rovibrational pumping (solid line).}
\label{fig:figure4}
\end{center}
\end{figure}

The rotational spectroscopy of the population in $v=0$ cannot be achieved with the pulsed dye laser alone. Its resolution ($0.1$  cm$^{-1}$) is not good enough to probe the rotational levels. Therefore, we use an additional narrow band cw DFB laser scanning the  B$^{1}\Uppi_{\rm u}(v_B=3,J_B)\leftarrow$X$^{1}\Upsigma_{\rm g}^{+}(v=0,J)$ transitions. After excitation and spontaneous emission, the rovibrational populations in $v=0$ are transferred to several vibrational levels, and mostly to those having a good Franck-Condon (FC) factor with the vibrational excited level \cite{2010PhRvL.105t3001A}. For our experiment, the FC factor between $v=7$ and $v_B=3$ is the largest available one ($\sim0.15$). The $v=7$ population is probed by the pulsed dye laser through the $9\leftarrow 7$ vibrational transition shown in Fig. \ref{fig:figure2}. Examples of rotational spectra obtained by scanning by the DFB laser diode frequency are given in Fig. \ref{fig:figure3} and \ref{fig:figure4}. In these figures, there is a systematic, irrelevant offset of $\sim7$ ions due to a measurement bias, and the real number of molecules in a specific rotational level is found by dividing the number of detected ions by the FC factor between $v_B=3$ and $v=7$ and the MCP efficiency ($\sim 30\%$).

In a first experiment, we perform a simple PA on $0^-_g (6s + 6p_{1/2})(v'=26,J')$. We then find vibrational spectra that hardly depend on the $J'$ value and look like the one shown in Fig. \ref{fig:figure2}(a). In accordance with the experimental and theoretical study related in Ref. \cite{2011PCCP...1318910L}, we find a distribution spread over more than 60 vibrational levels of the ground state. At this stage, as very few molecules are produced in $v=0$, we are unable to record any valuable rotational spectrum.

Next, by adding the femtosecond laser, we obtain the typical vibrational spectrum displayed in Fig. \ref{fig:figure2}(b) where we see that a substantial fraction ($\sim 25$\%) of molecules pumped into $v=0$ \cite{2011PCCP...1318910L}. The large population in $v=0$ enables us to study its rotational distribution. We find that rotational spectra depend on the $J'$ value: for $J'=1$, we find the upper spectrum shown in Fig. \ref{fig:figure3}, and for $J'=2$, the upper spectrum shown in Fig. \ref{fig:figure4}(a). In both cases, the rotational distributions are spread over less than four values of $J$, which indicates that vibrational optical pumping slightly broadens the rotational distribution \cite{2012_OptExpr_Bigelow_NaCs_Cooling}.  We see that when PA is tuned on the $J'=1$($2$) rotational level, only even (odd) $J$ values are populated. This is due to the parity conversion occurring at any stage of the all procedure.

The key result of this letter is obtained when rovibrational pumping is achieved. The corresponding vibrational spectrum in Fig. \ref{fig:figure2}(c) still shows an accumulation in $v=0$. However, with respect to the spectrum in Fig. \ref{fig:figure2}(b), we find more molecules in $v\neq0$. This is caused by the fact that the DFB laser has a slightly larger interaction zone than the femtosecond laser: some molecules, pumped out of $v=0$ by the DFB laser, remain in $v\neq 0$. The lower panel in Fig. \ref{fig:figure3} compares two rotational spectra obtained by rovibrational cooling and vibrational cooling in the case of a PA on $J'=1$. It undeniably indicates that the population of the absolute rotational ground state ($v=0,J=0$) has increased. The cooling results from an accumulation of absorption-spontaneous emission cycles using $J$ lowering absorption transitions, i.e. a laser spectrum covering mainly the $P$ branch transitions: B$^{1}\Uppi_{\rm u},(v_B=3,J-1)\leftarrow $X$^{1}\Upsigma_{\rm g}^{+}(v=0,2\leq J\leq 6)$. Due to the transition rules regarding the parity and the rotational quantum number, a single absorption-spontaneous emission cycle leads to either recover the same $J$ value or decrease it by two units. The frequency sweep, a sawtooth shape going from P(6) to P(2), may facilitate the cooling: if a molecule is excited from a given $J$ and immediately decays to $J-2$, it can be excited again by the same ramp. A few cycles are thus sufficient to transfer all the initial populations to $J=0$. Our procedure also affects the polarization of the molecular sample since it reduces the number of the $2J+1$ projections of the total angular momentum $\bm J$ on a given quantization axes. Indeed, the sample of unpolarized molecules is eventually transferred into a single well defined ($v=0,J=0,M_J=0$) state. This polarization is not straightforward because of the possible existence of dark states. That is why the two lasers used for vibrational and rotational cooling have different propagation axis and polarization states.

To illustrate the capabilities of the method, we also pump molecules toward other rotational levels. For instance, when PA is performed on $J'=2$, we obtain the results summarized in Fig. \ref{fig:figure4}(a): The accumulation occurs in the lowest odd-parity rotational level, i.e. ($v=0,J=1$).  Finally, Fig. \ref{fig:figure4}(b) shows the case where populations are optically pumped to a higher $J$ value, here $J=4$, by scanning the $R$ branch. 

The efficiency of these rotational pumping processes is defined as the number of molecules transferred into the intended rotational level divided by the number of molecules initially present in the other rotational levels when only vibrational pumping is applied. The results change a bit according to the type of pumping: respectively $\sim 40$\%, $\sim 30$\% and $\sim 25$\%  toward $J=0$, $J=1$ and $J=4$. In all the cases, a proportion of $\sim 30$\% remain in the other rotational levels. The residual part is lost in the $v\neq0$ levels as revealed by the comparison of Fig. \ref{fig:figure2}(b) and \ref{fig:figure2}(c). The weak number of photons involved in the process necessarily implies a negligible increase of temperature of $\sim1$ $\mu$K at most.

\begin{figure}[htbp]
\begin{center}
\includegraphics*[width=1\columnwidth]{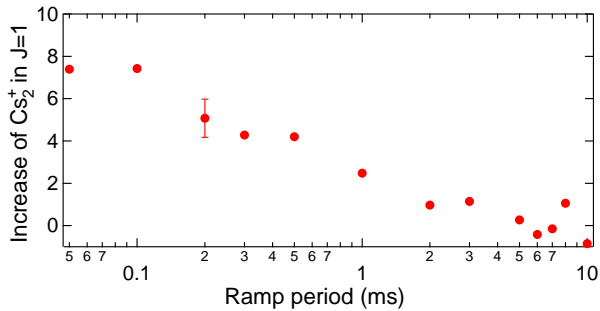}
\caption{(color online) Increase of the number of molecules in $J=1$ with its typical error bar versus the period of the frequency ramp applied on the narrow band laser used for rotational pumping. Because molecules escape from the interaction zone of the vibrational pumping laser, the transfer of molecules in $J=1$ decreases for a period longer than $100$ $\mu$s.}
\label{fig:figure5}
\end{center}
\end{figure}

To fully understand the mechanisms and limitations of our cooling method, we now detail the action of the pumping lasers. First, it is important to note that the vibrational cooling causes a little modification of $J$. For low $J$ values, typically only $1/5$ of the population ends up in an increased $J$ value after a single absorption-spontaneous emission cycle induced by the vibrational pumping laser. As the complete vibrational pumping to $v=0$ requires less than $5$ such cycles \cite{MatthieuViteau07112008}, about $50$\% of the molecules undergo an increase of $J$. We roughly estimate that it takes about $100$ $\mu$s to bring back molecules to $v=0$ from any $v$ when the saturation intensity of a typical rovibrational transition is  $\sim 100$ mW/cm$^2$ (considering typical FC factors of $0.1$ and Hönl-London factor of 1/4) and the vibrational cooling bandwidth is $\sim 100$ cm$^{-1}$. The minimum scan period ensuring a rotational pumping with a single ramp probability of $100$\% is also about $100$ $\mu$s, a time that is estimated through the maximum scan velocity roughly given by $\Gamma^2\sqrt{I/I_{sat}}$ where $\Gamma\approx 15\times 10^6$ s$^{-1}$ is the linewidth, $I$ the laser intensity, and $I_{sat}$ is the saturation intensity of some rovibrational transition. These considerations allow us to interpret the results in Fig. \ref{fig:figure5} where the relative transfer of molecules in $J=1$ is plotted versus the scan period. In principle, below $100$ $\mu$s, a single ramp does not allow a complete rotational pumping, but several ramps can be accumulated, which gives a pumping efficiency equivalent to that obtained with a $100$ $\mu$s scan period. On the contrary, once the scan period exceeds a few $100$ $\mu$s, the cooling efficiency decreases. This is explained by the molecular diffusion out of the laser interaction zone: Thus, the scan period limits the effective number of optical transitions. 

We have demonstrated that rovibrational pumping of molecules can be achieved by using different and complementary laser sources. Combined with the possibility of transferring populations from a given electronic state to another one \cite{2012_PRA_Horchani_Conversion}, we believe that such a method can be a very efficient tool to manipulate the rovibronic population of ultracold molecular samples or molecular beams. Besides, the efficiency of the rovibrational pumping should be easily enhanced by an optimized frequency shaping \cite{2010MolPh.108..795S} or with a DFB laser tuned on the $(v_B =0\leftarrow v=0)$ transitions rather than the $(v_B =3\leftarrow v=0)$ ones. Also, by using a more powerful femtosecond source, it becomes conceivable to realize an optical pumping of the external degrees of freedom based on a closed system made up of several states. In other words, the method can be a decisive help for (standard or Sisyphus) laser cooling of molecules \cite{2009PhRvA..80d1401Z,2009JPhB...42s5301R,shuman2010laser,2011_CRPhy_Perrin_laser}. 

Laboratoire Aimé Cotton is a member of Institut Francilien de Recherche sur les Atomes Froids (IFRAF) and of the LABEX PALM initiative. The exchange project between the University of Pisa and the University of Paris-Sud is acknowledged. A. F. and I. M. have been supported by the "Triangle de la Physique" under contracts No. 2007-n.74T and No. 2009-035T "GULFSTREAM" and No. 2010-097T-COCO2. We thank O. Dulieu and N. Bouloufa-Maafa for fruitful discussions as well as E. Dimova, L. Wang, L. Couturier, B. Le Crom and E. Mangaud for their help with the experiment.

\bibliographystyle{h-physrev}

\end{document}